\documentclass[12pt]{article}
\usepackage{psfrag}
\usepackage{amsmath,amssymb}
\usepackage{graphicx}
\usepackage{epsfig,rotate}
\usepackage{fancyheadings}

\newcommand{\vect}[1]{{\mathbf #1}}

\newcommand{\avg}[1]{\left\langle #1 \right\rangle}

\newcommand{\mum}{\ensuremath{\mu\text{m}}}

\begin{document}


\newcommand{\shorttitle}{Physics of Solutions and Networks of Semiflexible
  Macromolecules} \lhead[\sl\thepage]{\sl\shorttitle}
\rhead[\sl\shorttitle]{\sl\thepage} \cfoot{}

\renewcommand{\thefootnote}{\fnsymbol{footnote}}

\setcounter{page}{1}
\pagestyle{plain}

{\baselineskip=22pt
\noindent {\Large\sl Physics of Solutions and Networks of Semiflexible 
  Macromolecules and the Control of Cell Function}}
\footnote{\em Expanded
version of an invited talk given at the 14th Polymer Networks Group
International Conference in Trondheim (Norway), June 28 - July 3.  To be
published in The Wiley Polymer Networks Group Review Series, Volume 2}

\vspace{0.5cm}

\noindent{\large\sc Erwin Frey, Klaus Kroy and Jan Wilhelm}

\vspace{0.2cm}

\hrule

\tableofcontents

\section{Introduction}
\label{sec:Introduction}

Most of the concepts used to understand the viscoelastic properties of
chemical and physical gels of \emph{flexible} polymers require the
persistence length $\ell_{\text{p}}$ \cite{kratky-porod:49} to be
significantly smaller than other characteristic scales such as the
filament length $L$, the distance between crosslinks or the width of
reptation tubes. This condition no longer holds for networks of
\emph{semiflexible} polymers. One prominent family of such polymers
are cytoskeletal biopolymers like F-actin, intermediate filaments and
microtubules.  An impression of the typical conformations of the
filaments and the relative magnitude of the characteristic length scales can
be gained upon inspection of the following electron micrograph of a
semidilute actin solution.
\begin{figure}[hbt]
  \epsfxsize=0.4\columnwidth
  \epsffile{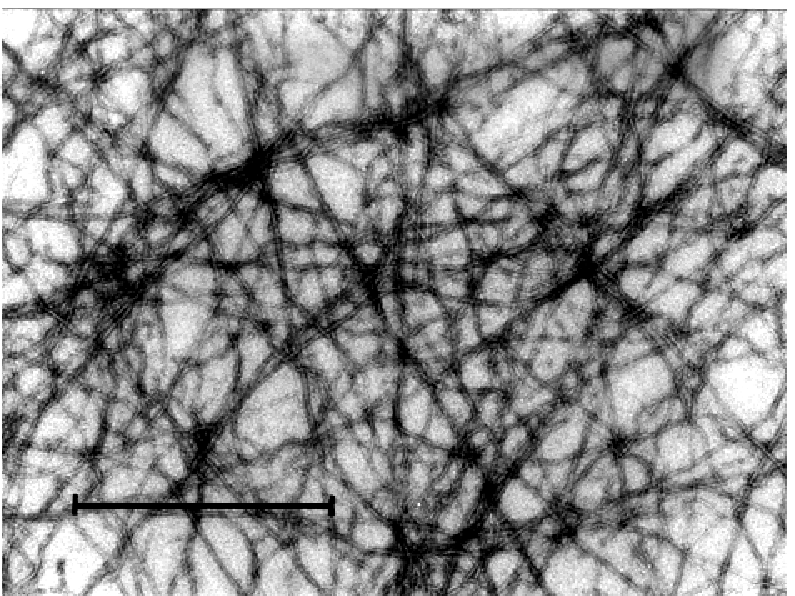} \hspace{1.5cm}
  \psfrag{r}{$\vect r(s)$} \psfrag{t}{$\vect t(s)$} \psfrag{s}{$s$} 
  \psfrag{0}{$0$} \psfrag{l}{$L$} 
  \epsfxsize=0.4\columnwidth
  \epsffile{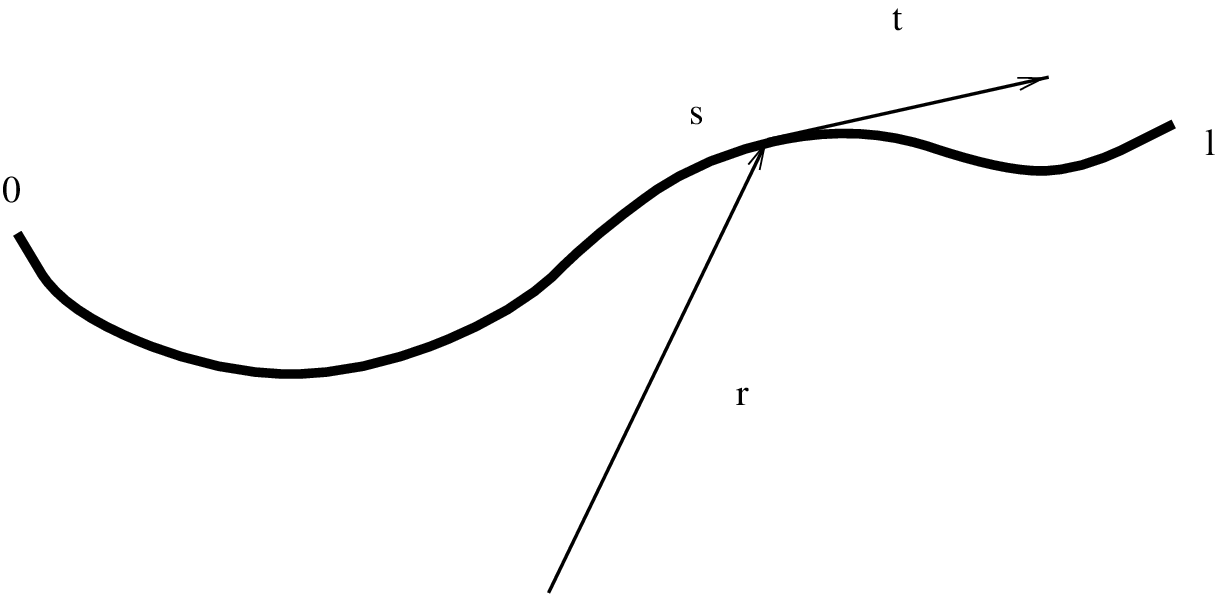}
  \caption{\small \emph{Left:} Electron micrograph of a 
    \protect $0.4 \, {\rm mg}/{\rm ml}$ actin solution.  The bar
    indicates 1 $\mu$m.  \emph{Right:} Sketch of the wormlike chain as
    a space curve ${\bf r} (s)$.}
\label{fig:micrograph}
\end{figure}
The most striking features of these networks are the enormous length and
relatively elongated structures of the constituent biopolymers.  Actin
filaments have a diameter of $7\,$nm \cite{holmes-etal:90} and can reach
lengths up to $30$-$100 \, \mu$m {\em in vitro}
\cite{burlacu-janmey-borejdo:92}, and several microns {\em in vivo}.  The
persistence length is approximately $17\,\mu$m
\cite{ott-magnasco-simon-libchaber:93,gittes-etal:93,isambert-etal:95}, quite
large compared to typical distances between neighboring filaments which is in
the range of a few tenth of a micron. This combination of length scales allows
biopolymers to form networks at very low volume fraction, so that solutions of
less than $0.1\%$ volume fraction of polymer are still strongly entangled. Thus
only a small amount of material needs to be produced by the cell in order to
generate a sufficiently strong network. This fact is not only of considerable
biological relevance but also facilitates interpretation of dynamic light
scattering experiments and their relation to theory
\cite{kroy-frey:97,kroy-frey:book}.

\pagestyle{fancy}

There are several motivations
\cite{janmey:91,luby_phelps:94,sackmann:94,mackintosh-janmey:97} for
studying the viscoelasticity of cytoskeletal networks
\cite{janmey-etal:88,mueller-etal:91,janmey-euteneuer-traub-schliwa:91,pollard-goldberg-schwarz:92,newman-etal:93,janmey-etal:94,kaes-etal:96,tempel-etal:96,hinner-etal:97,caspi-etal:98}:

(I) From a polymer physics perspective they are interesting because
their behavior is expected to be determined by principles and
mechanisms different from those established for flexible networks.  In
fact much of the physics behind the viscoelasticity of semiflexible
polymer networks is only being explored recently 
\cite{aharoni-edwards:94,jones-ball:91,%
   mackintosh-kaes-janmey:95,isambert-maggs:96,kroy-frey:96,%
   satcher-dewey:96,frey-kroy-wilhelm-sackmann:97,granek:97}.
 
 (II) From an experimental point of view, semiflexible polymer
 networks are interesting because several semiflexible polymers have
 persistence lengths on the order of several $\mum$ or even mm. Thus
 techniques such as optical microscopy of single fluorescence labeled
 filaments or attached colloidal probes can be used to study the
 behavior of the network at the \emph{single polymer level}
 \cite{janmey-etal:94,kaes-etal:96,caspi-etal:98,kaes-strey-sackmann:94}.
 For flexible polymers this has been possible only in simulations
 (see, e.g., \cite{binder-book:95}).
 
 (III) Finally, the viscoelastic properties and regulation of
 semiflexible polymer networks both inside cells and in the
 extracellular matrix are of significant importance for the mechanical
 stability and properties of biological tissue, for cell locomotion,
 adhesion and force generation
 \cite{janmey:91,luby_phelps:94,sackmann:94}.

\subsubsection*{The Cytoskeleton: Structure and Biological Role}
\label{sec:Cytoskeleton}

Living cells need both the ability to maintain their shape when exposed to
shear stresses exerted by their active contractile machinery or by fluid flow
in blood vessels and the ability to reorganize their shape and internal
architecture as is the case in cell migration and mitosis. The structure
responsible for the mechanical and dynamic properties of the cell is the {\em
  cytoskeleton}, a rigid yet flexible and {\em dynamic network of proteins} of
varying length and stiffness. Most eukaryotic cells contain three types of
protein filaments comprised of {\em actin}, {\em tubulin} and {\em intermediate
  filament proteins} such as vimentin. These, as well as the plasma-membrane
associated filaments make up the cytoskeleton \cite{schliwa:87}. There is also
a range of {\em accessory proteins} for each of the cytoskeletal filaments
which allow for {\em control} of nearly all mechanically relevant properties of
the protein filaments \cite{hartwig-kwiatkowski:91,olmsted:86}. Let us just
mention one example, gelsolin, which is frequently used in rheological
experiments. It caps the end of a growing actin filament and can thus be used
to regulate the average filament length in actin solutions.  There are many
other proteins with different tasks ranging from initiating and terminating
polymerization over introducing crosslinks and forming lateral arrays of
filaments to even changing the stiffness of the filaments.

An example of the biological role of the cytoskeleton is the migration
of an amoeba. Its motion is initiated by adhesion driven spreading of
the cell membrane on the substrate followed by gelation of actin in
the advancing lobe (pseudopodium). The cycle is completed by
retraction of the rear end and a gel-sol transition or fiber formation
in the advancing front \cite{sackmann:94}. This process is of course
quite complex and there is a subtle interplay between regulatory
mechanisms and the material properties of the cytoskeleton. But,
understanding the basic physical principles determining the
viscoelasticity of actin networks is certainly a prerequisite in
understanding such a biological process. In the following we will
address the following questions: (i) Can we even understand the
physics of a one-component system such as a purified F-actin solution
(depicted in Fig.~\ref{fig:micrograph})? (ii) Can we identify the
basic physical principles underlying the observed viscoelastic
behavior? (iii) How is it different from the physics of long flexible
coils?

\section{Single Chain Properties}

The model usually adopted for a theoretical description of
semiflexible chains is the {\em wormlike chain model}
\cite{kratky-porod:49,saito-takahashi-yunoki:67}. Here one describes
the filament as a smooth inextensible line ${\bf r} (s)$ of
length $L$ parameterized in terms of the arc length $s$.  The
statistical properties are determined by an effective free energy
(the ``Hamiltonian'')
\begin{equation}
\label{eq:hamiltonian}
  {\cal H}(\{\vect r(s)\}) = \frac{\kappa}{2} \, \int_0^L \!\! ds 
             \left( \frac{\partial^2 {\bf r} (s) }{\partial s^2} \right)^2 \, ,
\end{equation}
which measures the total elastic energy of a particular conformation by the
integral over the square of the local curvature weighted by the \emph{bending
  modulus} $\kappa$. The inextensibilty of the chain is expressed by the local
constraint, $|{\bf t}(s)|=1$, on the tangent vector ${\bf t} (s) = \partial
{\bf r} / \partial s$.  We will see that this constraint is essential for a
correct description of the static as well as the dynamic properties of
semiflexible polymers.  Due to the mathematical complications resulting from
the inextensibilty only few of the statistical properties of the wormlike chain
can be extracted analytically, the best known being the exponential decay of
the tangent-tangent correlation function $ \langle {\bf t} (s) {\bf t} (s')
\rangle = \exp \left( -|s-s'| / \ell_p \right)$ with the persistence length
$\ell_p = \kappa / k_B T$, the mean-square end-to-end distance
\cite{kratky-porod:49} $ {\cal R}^2 := \langle \left[ {\bf r }(L)-{\bf r}(0)
\right]^2 \rangle = L^2 f_D (L/\ell_p)$, and the radius of gyration
\cite{benoit-doty:53} $ {\cal R}_g^2 = \ell_p^2\left(f_D(L/\ell_p)-1+
  L/3\ell_p\right)$, where $f_D (x) := 2(e^{-x}-1+x)/x^2$ is the Debye
function.

\subsection{Force-Extension Relation}

\begin{figure}[htb]  
\centerline{\epsfxsize=0.23\textwidth 
\rotate[r]{\epsffile{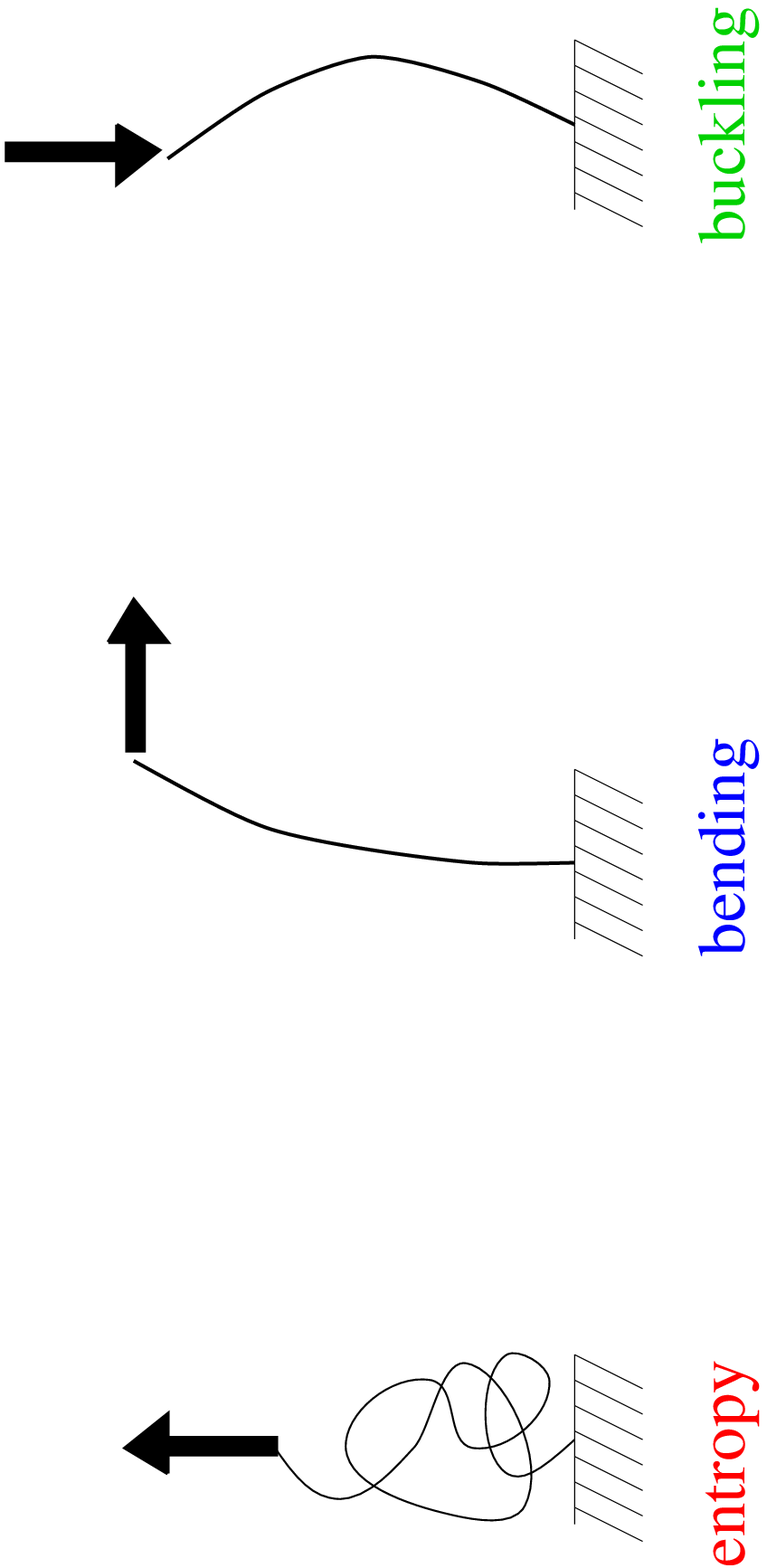}}
\hspace{1cm}\includegraphics[width=0.42\textwidth]{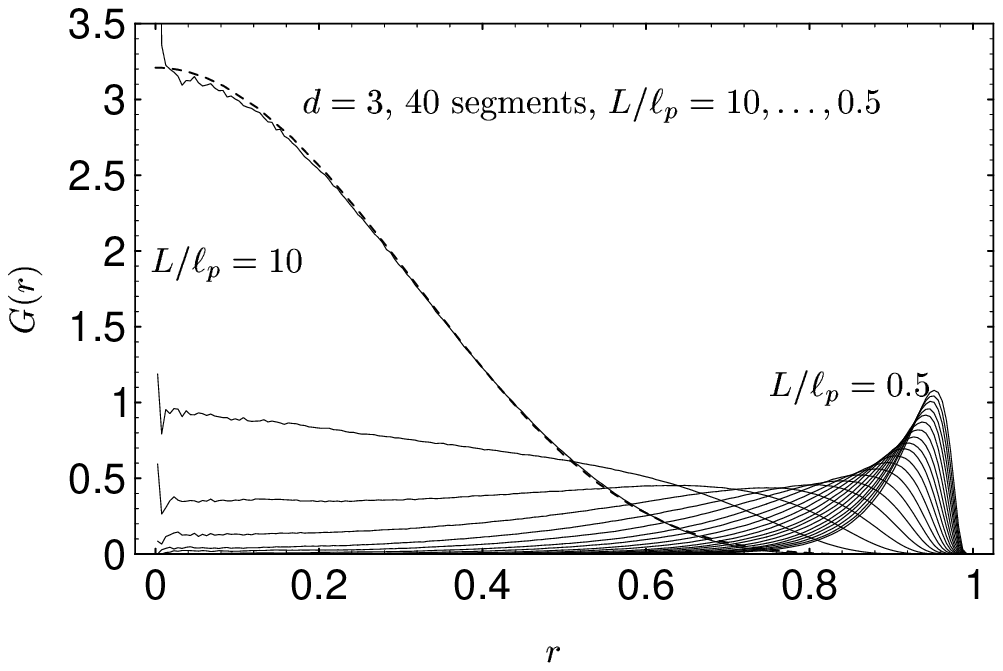}}
\caption{\label{fig2}
  \small \emph{Left:} A flexible chain's linear response is purely entropic and
  isotropic. The elastic response of a stiff rod is extremely anisotropic.
  Transverse forces (bending) lead to a purely mechanical response, whereas the
  longitudinal response (buckling) is characterized by a spring constant
  inversely proportional to the temperature. \emph{Right:} Numerical results
  for the end-to-end distribution function of a (discretized) wormlike chain in
  $d=3$ dimensional space (numerical data from Ref.\ 
  \protect\cite{wilhelm-frey:96}).  With increasing stiffness there is a
  pronounced crossover from a Gaussian shape to a form with the weight shifting
  towards full stretching. The dashed line indicates the Daniels approximation
  \protect\cite{daniels:52}.}
\end{figure}

One of the most obvious differences between flexible and semiflexible
polymers is their response to external forces (see Fig.\ 
\ref{fig2}~(left)). In the flexible case the response
is isotropic and proportional to $1/k_B T$, i.e., the Hookian force
coefficient is proportional to the temperature.  When the persistence
length is of the same order of magnitude as the contour length, the
response becomes increasingly {\em anisotropic}.

Then the linear response of the chain depends on the orientation of
the force with respect to the tangent vector at the clamped end.
Transverse forces give rise to ordinary mechanical bending of the
filaments and the {\em transverse spring coefficient} is proportional
to $\kappa$.  The linear response for longitudinal
forces is due to the presence of thermal undulations, which tilt parts
of the polymer contour with respect to the force direction.  The
effective {\em longitudinal spring coefficient} turns out to be
proportional to $\kappa^2/T$ indicating the breakdown of linear
response at low temperatures ($T\to0$) or very stiff filaments
($\ell_p\to\infty$).  This is a consequence of the Euler buckling
instability. Note also that for the special boundary conditions of a
grafted chain (as depicted in Fig.~\ref{fig2}~(left))
the linear response of the chain can even be worked out exactly for
arbitrary stiffness \cite{kroy-frey:96}; these calculations use the
fact that the conformational statistics of the wormlike chain is
equivalent to the diffusion on the unit sphere
\cite{saito-takahashi-yunoki:67}.

\subsection{Radial Distribution Function}

An important quantity describing the statistical properties of the chain is the
\emph{probability distribution of the end-to-end vector} $G(\vect r;L) =
\avg{\delta(\vect r-\vect R)}$.  For a \emph{freely jointed phantom chain} this
function is known exactly \cite{yamakawa}. As for any model with short-ranged
interactions it converges quickly to a Gaussian distribution $G_0(\vect r;L)
\sim \exp \left( - {3 r^2}/{4 \ell_pL} \right)$ for an increasing number of
segments.  For chains that are at least some $10\,\ell_p$ long the Gaussian can
serve as an excellent approximation to $G(\vect r;L)$ for many purposes. For
the freely jointed chain the persistence length $\ell_p$ is independent of
temperature because its microscopic origin lies in steric constraints rather
than in the bending stiffness of the backbone.

For a Gaussian chain, the separation by a given distance $r$ of any two
segments with preferred mean-square distance $2\ell_ps$ is punished by the free
energy cost $F ({\bf r}) = - k_B T \ln G_0({\bf r};s) = const. + 3 k_B T {\bf
  r}^2 / 4\ell_ps$ {\em quadratic} in the end-to-end distance. Due to the Euler
instability this is very different for semiflexible chains. The characteristic
feature of the physics of beam buckling is that the energy $E_{\text{cl}}$ of a
straight rod is an almost {\em linear} function of its end-to-end distance $R$,
$ E_{\text{cl}} \approx f_c \cdot (L-R) $.  Here $f_c = \kappa \pi^2/L^2$ is
the critical force for the onset of the Euler instability. Neglecting
fluctuations around the classical contour this would lead to an end-to-end
distribution function with maximum weight at $R=L$, $G({\bf r};L) \propto \exp
[- f_c\cdot (L-r)/ k_B T]$.  Note that with such an approach we completely
ignore entropic effects which are the only contributions in case of the freely
jointed chain, discussed above. In order to correct for this omission we have
to multiply the above Boltzmann weight by the relative number of allowed
conformations.  This becomes most obvious for a completely stretched chain,
where up to global rotations only one possible configuration exists and
consequently the end-to-end distribution function has to vanish. These
qualitative arguments lead to the shape of the distribution function shown in
Fig.\ \ref{fig2}~(right).  The actual form of the end-to-end distribution
function can be obtained within a quantitative analysis \cite{wilhelm-frey:96}
of the wormlike chain.

\subsection{Dynamic Light Scattering}

A useful experimental technique for investigating the short time dynamics of
semiflexible polymers is dynamic light scattering (DLS).  In DLS experiments
one directly observes the dynamic structure factor $g({\bf q},t)$. We focus on
the ideal case of a dilute or semidilute solution of semiflexible polymers,
where the scattering wavelength is much smaller than the mesh size. We also
assume a separation of length scales, $a \ll \lambda \leq \ell_p,L$, i.e., the
scattering wavelength $\lambda$ is large compared to the monomer size $a$ but
small compared to the characteristic mesoscopic scale defined by $L$ and
$\ell_p$.  As a consequence the contributions to the time decay of $g({\bf
  q},t)$ from center of mass and rotational degrees of freedom of the chain are
strongly suppressed as compared to contributions from bending undulations.
Moreover, for this case \cite{frey-nelson:91,farge-maggs:93,%
  harnau-winkler-reineker:96,kroy-frey:97,granek:97} the structure
factor can be written as $\exp(-q^2r^2_\perp(t)/4)$ with the local
mean square displacement $r^2_\perp(t) \sim t^{3/4}$:
\begin{equation}
\label{stretched_exponential}
g({\bf q},t)\propto \exp[-(\gamma_q t)^{3/4}] \, ,
\end{equation}
where $\gamma_q \sim q^{8/3}/\zeta_\perp \ell_p^{1/3}$
\cite{kroy-frey:97,kroy-frey:book}.  Such a stretched exponential
behavior has been confirmed experimentally with very high accuracy for
F-actin solutions \cite{goetter-etal:96}. However, a more careful
analysis reveals that it cannot hold for very short times.
\begin{figure}[htb]
  \centerline{\epsfxsize=0.415 \columnwidth\epsffile{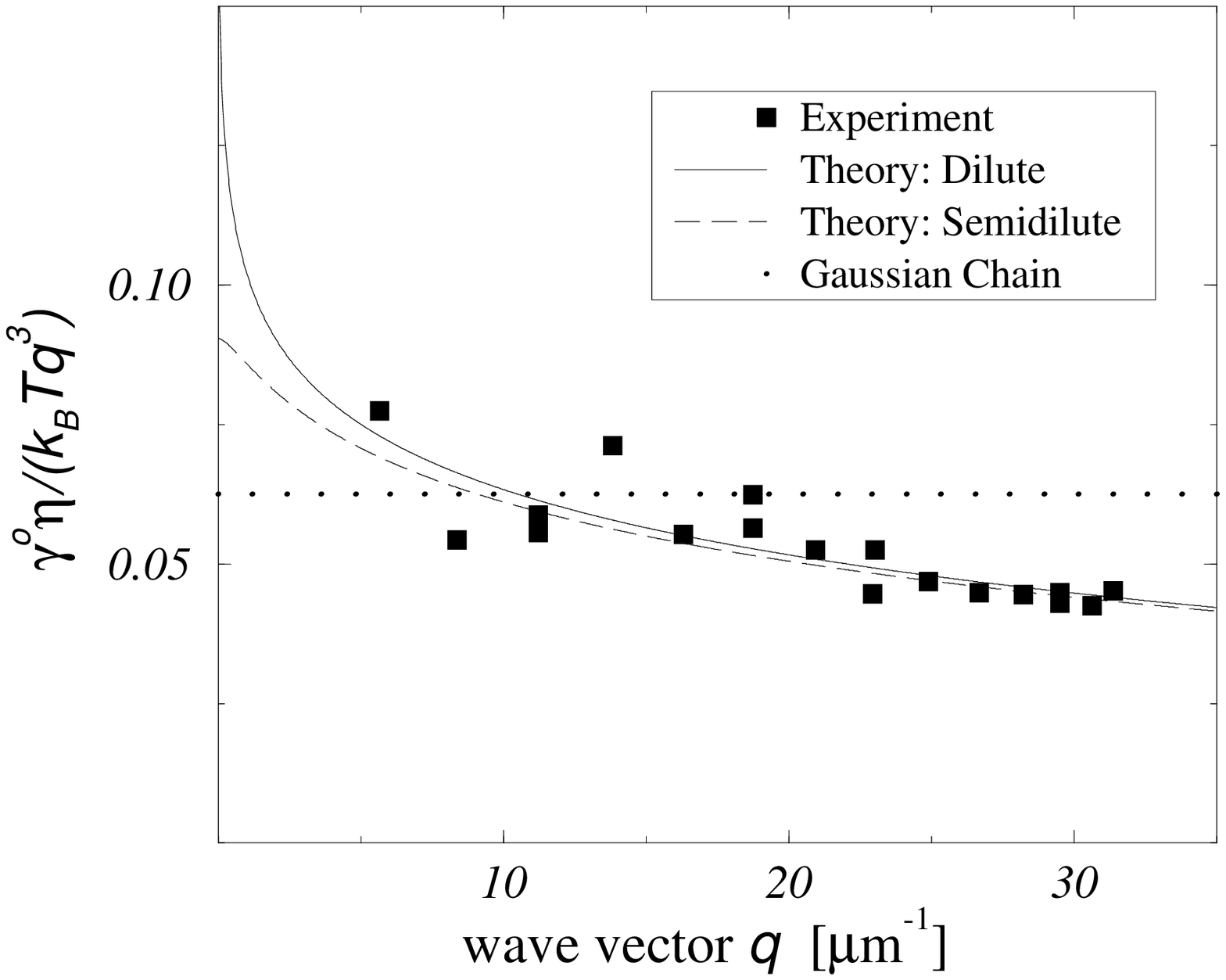}
    \hspace{1cm} \epsfxsize=0.45\columnwidth \epsffile{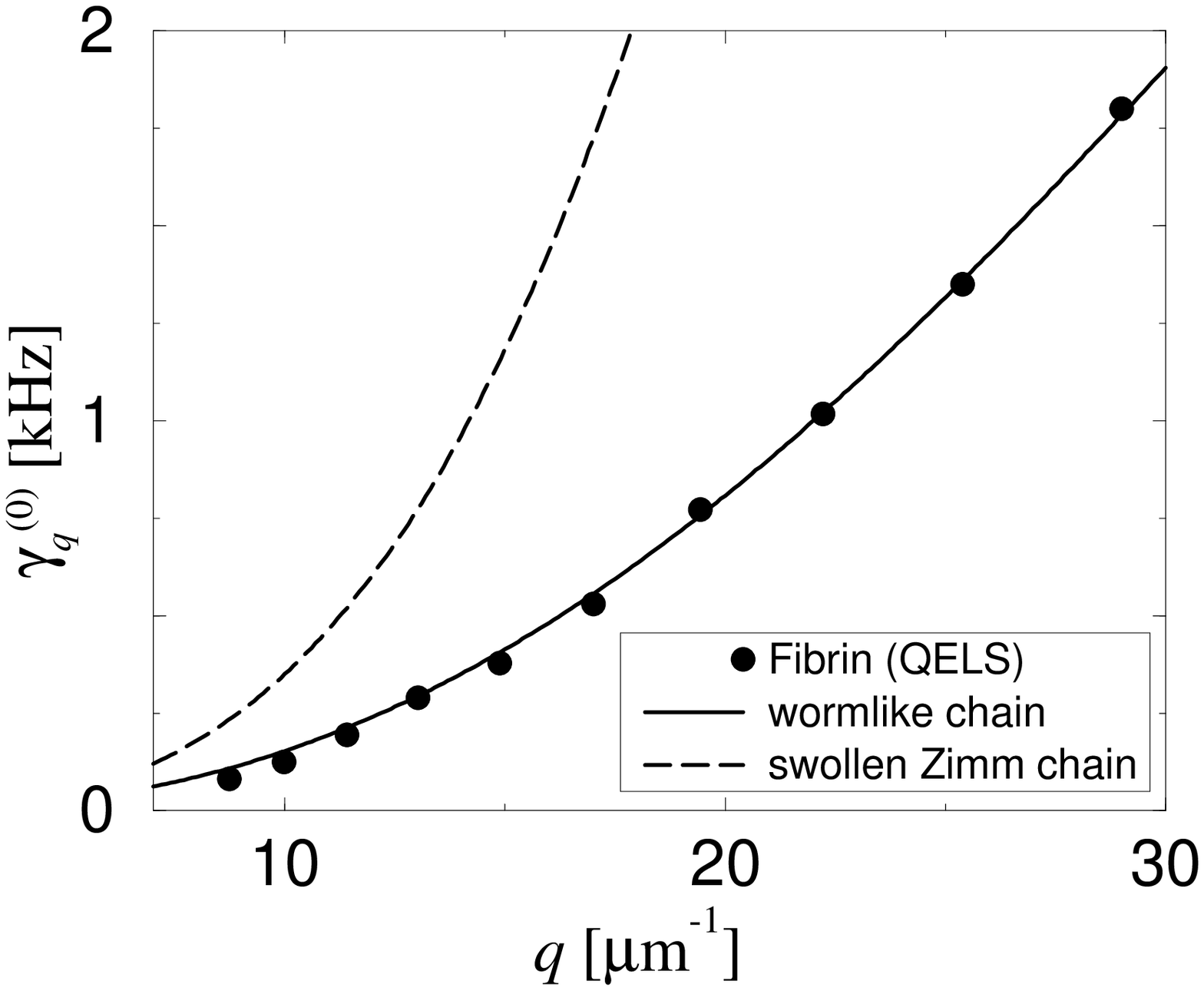}}
\caption{\small  \emph{Left:} The $q-$dependence of the initial
  decay rate compared to DLS data on actin \protect \cite{schmidt:phd}.
  \emph{Right:} Comparison of the classical result for a swollen Zimm chain
  \protect\cite{doi-edwards:86} with Eq.~(\protect\ref{gamma_0}) and
  quasi-elastic light scattering experiments with the semiflexible biopolymer
  fibrin \protect\cite{arcovito-etal:97}. The data were most kindly provided by
  G.  Arcovito (see also this volume).}
\label{fig:arcovito}
\end{figure}
For times shorter than $\zeta_\perp/\kappa q^4$ the
bending forces can be considered weak and the contour obeys the fast wiggling
motion imposed by hydrodynamic fluctuations.  As a consequence the initial
decay is of the form $g({\bf q},t)\propto
\exp(-\gamma_q^{(0)}t)$ with \cite{kroy-frey:97}
\begin{equation}\label{gamma_0}
  \gamma_q^{(0)}= \frac{2k_BT}{3\pi \zeta_\perp}q^3 = 
\frac{k_BT}{6\pi^2 \eta}q^3\ln \left( e^{5/6}/ka\right).
\end{equation}
For polymers, which are not quite as stiff as actin, e.g.\ for so called
intermediate filaments, this initial decay regime is readily observed in light
scattering experiments. A very convincing confirmation of these theoretical
results has recently been found in fibrin systems \cite{arcovito-etal:97} (see
Fig.~\ref{fig:arcovito}).  Analyzing the data by Eq.~\ref{gamma_0} allows one
to estimate the friction coefficient $\zeta_\perp$ entering the Langevin
equation or, equivalently, the thickness $a$ of these filaments.

\section{Collective Properties}

In conventional polymer systems made up of long flexible chain molecules the
viscoelastic response is {\em entropic} in origin over a wide range of
frequencies \cite{doi-edwards:86}. For semiflexible polymers a complete
understanding of the viscoelastic response is complicated by several factors.
First of all, there are several ways by which forces can be transmitted in a
network. This can either happen by steric (or solvent-mediated) interactions
between the filaments or by viscous couplings between the filaments and the
solvent undergoing shear flow. It is a priori not at all obvious which if any
of these coupling will dominate. In the case of flexible polymers it is
generally believed that macroscopic stresses are transmitted in such a way that
these transformations stay affine locally, i.e.\ that the end-to-end distance
of a single filament follows the macroscopic shear deformation
\cite{doi-edwards:86}.  Second, single filaments are {\em anisotropic elastic
  elements} showing quite different response for forces perpendicular or
parallel to its mean contour. Therefore one has to ask what kind of deformation
of the actin filament is the dominant one and whether due to the anisotropy of
the building blocks of the network macroscopically affine deformations stay
affine locally. In the following we will address some of the issues raised.

\subsection{Plateau Modulus for Entangled Solutions}

If solutions of semiflexible polymers are sufficiently dense and are probed on
sufficiently short time scales (typically in the range of $10^{-2}$~Hz to
$1$~Hz) they will exhibit a {\em ``rubber plateau''}. Its existence is in
general traced back to a {\em time scale separation} between the internal
dynamics and the center of mass motion of the polymers. An externally imposed
shear stress will then be transmitted to the individual strands, whose response
will determine the magnitude of the modulus. This many chain problem is usually
reduced to a single chain model by making certain assumptions on the effect of
the mutual steric constraints on the conformation of a single filament.

In what might be called the {\em affine model} the ``phantom model''
\cite{treloar:book} is adopted to semiflexible polymer systems
\cite{mackintosh-kaes-janmey:95}. It is assumed that upon deforming the network
macroscopically the path of a semiflexible polymer between two entanglement
points is straightened out or shortened in an affine way with the sample. The
macroscopic modulus is then calculated from the free energy cost associated
with the resulting change in the end-to-end distance. Since in a solution
forces between neighboring polymers can only be transmitted transverse to the
polymer axis and there is no restoring force for sliding of one filament past
another, it is however hard to imagine that entanglements are able to support
longitudinal stresses in filaments. The modulus predicted in the affine model
should scale as $G^0 \propto c^{11/5}$ and leads to absolute values of the
order of $10$~Pa; such high values are at odds with the low values observed in
recent experiments on F-actin solutions \cite{tempel-etal:97}. It was therefore
argued \cite{mackintosh-janmey:97} that such models are more appropriate for
crosslinked networks, where they would predict a plateau value $G^0 \simeq k_B
T \ell_p^2 / \xi_m^5$.  But, even in such chemical networks with crosslinks
present it is a priori not obvious that local deformations on the scale of a
single filament are actually affine and that longitudinal stresses in the
filaments are the dominant contribution to the plateau modulus (see also
section \ref{sec:crosslinking}).

\begin{figure}[htb]
  \begin{center}
    \includegraphics[width=0.5\columnwidth]{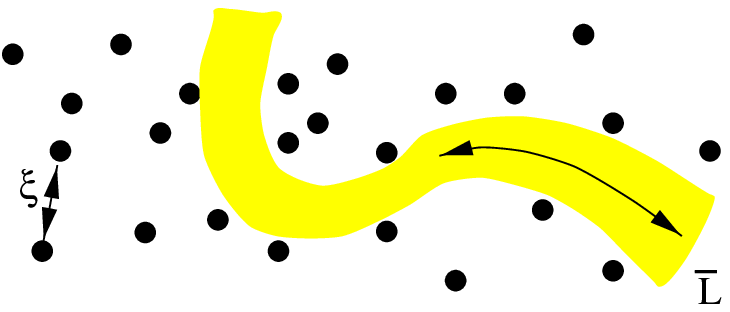}
    \hspace{0.5cm}
     \includegraphics[width=0.45\columnwidth]{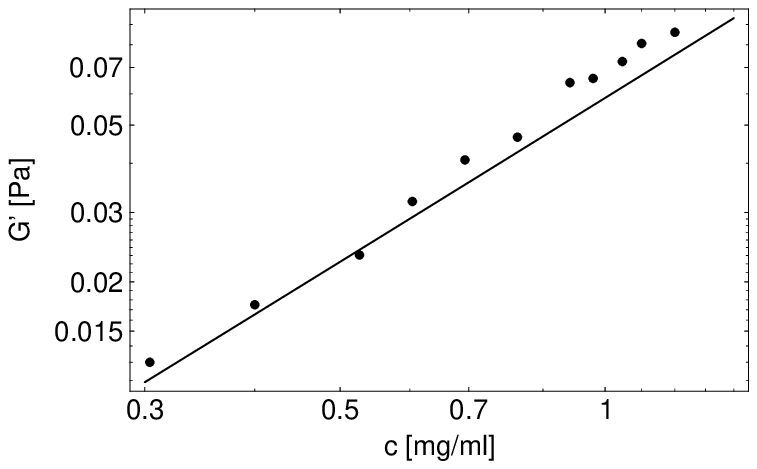}
    \caption{\small \emph{Left:} A semiflexible polymer
      can trade bending energy for a wider tube.  The configuration of the
      constraining polymers (dots) is the same as in the upper figure.
      \emph{Right:} Comparison of the predicted shear modulus
      \protect\cite{wilhelm-frey:98} to experiment
      \protect\cite{hinner-etal:97}. The total length $L$ and persistence
      length $\ell_{\text{p}}$ were set to $\ell_{\text{p}} = 17\,\mum$ and $L
      = 16\,\mum$, respectively \protect\cite{hinner-etal:97}.}
    \label{fig:tube_tempel}
  \end{center}
\end{figure}

Recent theoretical and experimental studies
\cite{isambert-maggs:96,tempel-etal:97} based on
Refs.~\cite{helfrich-harbich:85,odijk:86,semenov:86} suggest a different view.
Here one considers the free energy cost of suppressed transverse fluctuations
of the polymers that comes about by an {\em affine deformation of the tube}
diameter.  According to Odijk \cite{odijk:86} the mean distance between
collisions of a tagged polymer with its surrounding tube with diameter $d$ is
given by the deflection length $L_e \simeq \ell_p^{1/3} d^{2/3}$.  Since each
of these collisions reduces the conformation space there is a free energy of
the order of $k_B T$.  The total free energy of $\nu = c/L$ polymers per unit
volume becomes $ F \simeq \nu \, k_B T \, {L}/{L_e}$.  To be able to compare
these results to experiments one needs to know how the tube diameter $d$
depends on the concentration of the solution or equivalently on the mesh size
$\xi_m := \sqrt{3/\nu L}$. In other words we have to determine the average
thickness $d$ of a bend cylindrical tube in a random array of polymers as
depicted in Fig.~\ref{fig:tube_tempel}~(left).

The contour and thickness of the tube will be determined by a competition
between bending energy favoring a thin straight tube and entropy favoring a
curved thick tube. This competing effects define a characteristic length scale
which turns out to be $L_e$. For length scales below $L_e$ the tube will be
almost straight and we can estimate its thickness as follows. Upon restricting
the orientations of the polymers to being parallel to the coordinate axes the
density of intersection points (black dots in
Fig.~\ref{fig:tube_tempel})~(left) will be $1/\xi^2$. Hence for a tube of
length $L_e$ the line density of these intersection points projected to a line
perpendicular to the tube increases as $L_e / \xi_m^2$ which implies that the
tube diameter decreases with increasing tube length as $d \simeq
{\xi_m^2}/{L_e}$.  Hence one finds $L_e = (\xi_m^2 \ell_p^{1/2})^{2/5}$ leading
to the following form of the free energy and hence the plateau modulus
\begin{eqnarray}
 G^0 \simeq F \simeq  k_B T \, \ell_p^{-1/5} \, c^{7/5} \, .
\label{tube_modulus}
\end{eqnarray}
The above scaling law is included as a limiting case in a more detailed
analysis concerned with the calculation of the absolute value of the plateau
modulus \cite{wilhelm-frey:98}. The same scaling result has been obtained
previously \cite{isambert-maggs:96} using a different scaling argument.

Recent experiments seem to favor the above tube picture, where the plateau
modulus is thought to arise from free energy costs associated with deformed
tubes due to macroscopic stresses.  Fig.~\ref{fig:tube_tempel}~(right) shows
the results of a recent measurement of the concentration dependence of the
plateau modulus in F-actin solutions \cite{tempel-etal:97} which agrees well
with the scaling prediction $G^0 \propto c^{7/5}$.

\subsection{Viscoelasticity and High Frequency Behavior}

At frequencies above the ``rubber plateau'' (i.e.\ above $1$~Hz for a
typical F-actin solution) a power-law increase of the storage and loss
modulus with frequency, $G'(\omega) \propto G''(\omega) \propto
\omega^{3/4}$, has been observed
\cite{amblard-etal:96,gittes-etal:97,schnurr-etal:97}. It is tempting
to speculate that it is somehow tightly connected with the anomalous
subdiffusive behavior of the segment dynamics of a single filament.
But in view of the actual micro-rheological experiments, where one
observes the mean-square displacement of a bead of diameter larger
than the mesh size and hence couples to a large number of filaments,
it is not obvious how this comes about. A thorough understanding would
need to explore the nature of the crossover from local dynamics
dominated by filament undulations to the collective dynamics of the
network and the solvent.

At present there are two different theoretical approaches based on different
assumptions on the nature of the dominant excitations of the individual
filaments generated by the beads embedded in the network.  In one class of
theoretical models one takes over the above mentioned ``phantom model'' to the
high frequency behavior \cite{gittes-mackintosh:98,morse:98a}. It is assumed
that under an applied shear deformation the filaments undergo affine
deformations on a length scale of order $L_e$ implying longitudinal stresses on
single filaments.  In the high frequency regime this leads to
\cite{gittes-mackintosh:98,morse:98a}
\begin{eqnarray}
  G^*(\omega) = 
  \frac{1}{15} \nu (k_B T)^{1/4} \ell_p^{5/4} (i \omega \zeta_\perp)^{3/4} \, ,
\label{hf_modul_long}
\end{eqnarray}
independent of the entanglement length $L_e$.
 
A complementary theoretical approach \cite{kroy-frey:unpub} starts from an
effective medium description for the polymer solution at large scales which
crosses over to the single polymer picture at about the tube diameter. The low
frequency response is due to peristaltic modes of the effective medium. At high
frequencies, the penetration depth for these modes falls below the tube
diameter and the excitations are bound to the polymer backbones. Assuming that
the forces between polymers are transmitted by binary collisions, the
transverse modes that make up the plateau modulus according to the tube model,
are also responsible for the high frequency response.  This again leads to an
$\omega^{3/4}-$asymptotics of $G^*(\omega)$ at high frequencies. However, the
model describes the crossover to and the moduli within the plateau region and
allows scaling predictions for the relationship between plateau modulus and
entanglement frequency. The information contained in the viscoelastic moduli is
conveniently expressed in terms of the density of relaxation modes. Preliminary
investigations show that already the simplest scaling assumption for this
density (which certainly greatly over-simplifies the complicated crossover from
single polymer dynamics to the effective medium modes) leads to excellent
agreement with experimental data. Using the fluctuation-dissipation theorem,
the long time behavior of the dynamic structure factor in semidilute solutions
can also be derived.

Which one of these theoretical models is capturing the correct physics is not
clear at present.  It may well be that the actual physical mechanism is
different from both.  There is certainly a tremendous need for more detailed
experimental studies which not only measure the power-law dependence of the
modulus but also determine the concentration dependence of the entanglement
frequency.

\subsection{Effect of crosslinking} 
\label{sec:crosslinking}
For a crosslinked network of semiflexible polymers bending \emph{and}
compressing forces can be transmitted to the filaments. Both for networks where
the mesh size is very small compared to the persistence length so that the
longitudinal elastic response of the polymers is dominated by their Young's
modulus and for networks with larger mesh size where thermal undulations are
crucial in understanding the elastic response of single filaments
\cite{mackintosh-kaes-janmey:95,kroy-frey:96}, compression is a much stiffer
mode of deformation than bending. Unless highly ordered network geometries are
assumed, it is not clear which of the two modes will dominate the elastic
response.  Different assumptions on the real or effective network geometry can
either favor the bending modes as in \cite{satcher-dewey:96} or the
compressional modes as in \cite{mackintosh-kaes-janmey:95} leading to
substantially different predictions for the modulus.

We used a two-dimensional toy model to investigate which type of deformation
mode is dominant in a disordered crosslinked network. Sticks of length $L$ were
placed randomly on the plane and crosslinked at every intersection with another
stick. Crosslinks were inextensible. Sticks were assigned a Young's modulus $E$
and a diameter $r$ resulting in force constants $k_{\text{comp}} = \pi r^2 E$
for compression and $k_{\text{bend}} = 3 \pi r^4 E/3 L^2 = 3 \kappa/L^2$ for
bending the rod with one end clamped. Units were chosen such that $L = 1$ and
$\kappa = 1$. The model was subjected to periodic boundary conditions, strained
and the linear elastic response calculated by the method of finite elements.
While this is a purely mechanical model it captures the essential features of
two very different force constants and disorder. Entropic contributions from
fluctuations of the crosslink positions are not expected to be significant for
dense networks.

Which of the two modes dominates the elastic behavior was determined by keeping
$k_{\text{bend}}$ fixed and varying $k_{\text{comp}}$.
\begin{figure}[htb]
\centerline{\includegraphics[height=.35\textwidth]{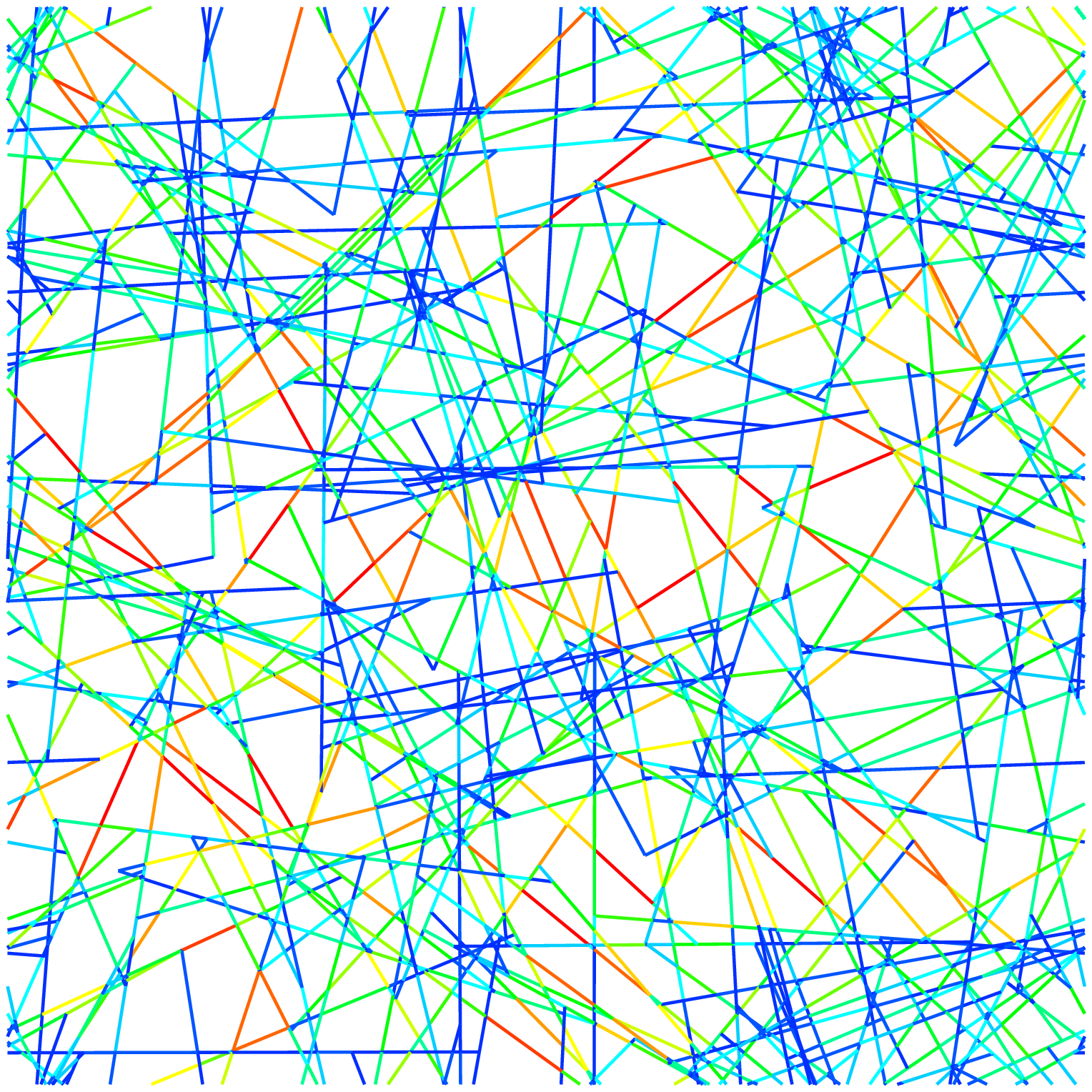}
  \hspace{1cm} \includegraphics[height=.35\textwidth]{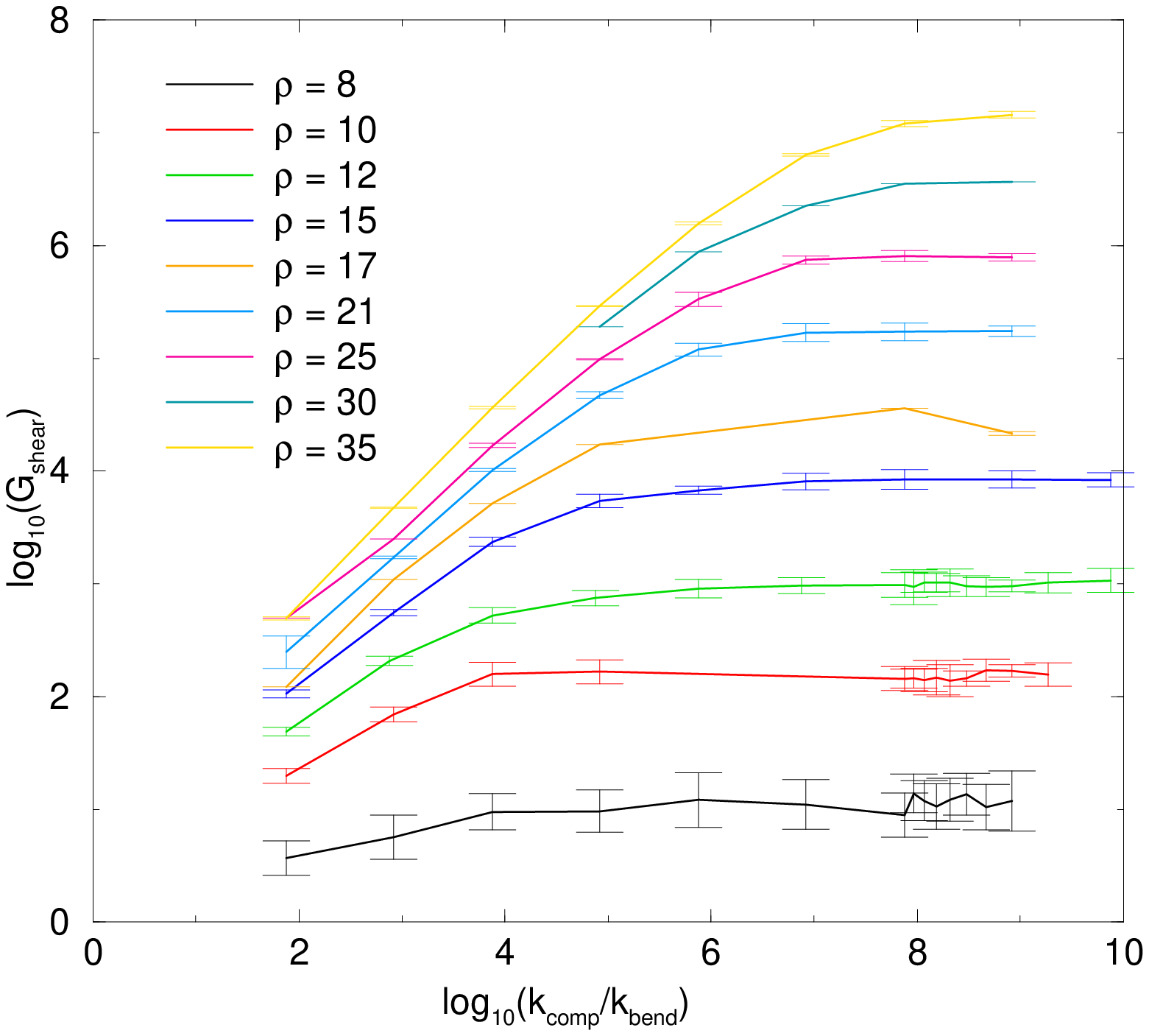}}
\caption{\label{fig:crosslinked_network}
  \small {\it Left}: Network of sticks for $\rho = 50$, $L = 2$ and
  $\alpha = 0.01$. The color code indicates the load distribution with
  energy decreasing from red to blue. {\it Right:} Dependence of the
  shear modulus on the ratio $k_{\rm comp}/k_T/k_{\rm bend}$ for
  networks with $L = 15$.}
\end{figure}
We observe that for slender rods or low densities a certain point the modulus
ceases to depend on $k_{\rm comp}$, indicating that the elasticity is dominated
by bending modes.  While these two-dimensional results are certainly not
straightforwardly applicable to three-dimensional networks we will nevertheless
try to get a feeling for the scales involved. Network densities can be compared
roughly by using the average distance $L_c$ between intersections as a measure:
A cytoskeletal network might have $L_c \approx 0.1\,\mu\text{m}$ with typical
filament lengths of $2\,\mu\text{m}$ corresponding to a two-dimensional density
of $\rho \approx 20$ and an aspect ratio of $\alpha\approx 0.002$ resp.
$k_{\rm comp}/k_{\rm bend} \approx 10^{5}$. Comparison with
Fig.~\ref{fig:crosslinked_network} shows that this would just place the network
in the bending dominated regime.  This might, however, be different for
different scales or if more order is present in the network than assumed here.
For a more detailed analysis of the random stick model see
Ref.~\cite{wilhelm-frey:97b}.

\section{Summary and Future Perspectives}

We have seen that the cytoskeleton is a composite biomaterial with a wide
variety of interesting viscoelastic properties. In particular F-actin solutions
and networks provide a model system for a polymeric liquid composed of
semiflexible polymers which is accessible to a complementary set of
experimental techniques ranging from direct imaging techniques over dynamic
light scattering to classical rheological methods. From these studies it has
become quite obvious that semiflexible polymer networks require new
theoretical models different from conventional theories for rubber elasticity.
The nature of the entanglement in solutions of filaments is very different from
flexible coils. In a frequency window where an elastic plateau is observed a
tube picture where the modulus results from the free energy costs associated
with the tube deformations seems to be sufficient to explain the observed
concentration dependence of the plateau modulus and even its absolute value
\cite{wilhelm-frey:98}.

Outside the rubber plateau in the high-frequency as well as the low-frequency
regime the situation is less clear.  Micro-rheology and dynamic light
scattering experiments allow us to access the short-time dynamics of the
filaments within a network.  Here a theoretical model which describes the
combined dynamics of network and solvent in this regime is still lacking. At
present there are two quite different approaches which either start from a
continuum medium approximation or from a single-filament picture. Obviously
both are just limiting cases and a molecular theory needs to explain how
starting from the single-filament dynamics including interactions with the
solvent and the neighboring filaments leads at some length and time scale to
collective behavior, which might be described in terms of some continuum model.

Another very important question is concerned with the effect of chemical
crosslinks on the mechanical properties of semiflexible polymer networks. This
is of prime interest for both cell biology and for polymer science.  In cell
biology one would like to know how the material properties (e.g.\ elastic
modulus, time scales for structural rearrangement and stress propagation)
change with the network architecture and the mechanical and dynamic properties
of the crosslinks. From the perspective of polymer science it connects
cytoskeletal elasticity with the very active fields of transport in random
media and elastic percolation. In section \ref{sec:crosslinking} we have
presented a numerical study using a two-dimensional toy model.  One can
certainly not expect that such a simplified model leads to quantitative
results, but we think that some of its main features carry over to the more
complicated situation of a three-dimensional network. \\
{\small {\bf Acknowledgment:} This work has been supported by the Deutsche
  Forschungsgemeinschaft through a Heisenberg Fellowship (No.\ Fr\ 850/3) and
  through SFB\ 266 and 413.}

\end{document}